\documentclass[english,prd,superscriptaddress,nofootinbib,preprintnumbers]{revtex4}
\usepackage[T1]{fontenc}
\usepackage[latin9]{inputenc}
\usepackage{bm}
\usepackage{amsmath}
\usepackage{amssymb}
\usepackage{esint}

\makeatletter
\@ifundefined{textcolor}{}
{%
 \definecolor{BLACK}{gray}{0}
 \definecolor{WHITE}{gray}{1}
 \definecolor{RED}{rgb}{1,0,0}
 \definecolor{GREEN}{rgb}{0,1,0}
 \definecolor{BLUE}{rgb}{0,0,1}
 \definecolor{CYAN}{cmyk}{1,0,0,0}
 \definecolor{MAGENTA}{cmyk}{0,1,0,0}
 \definecolor{YELLOW}{cmyk}{0,0,1,0}
}

\@ifundefined{definecolor}
 {\usepackage{color}}{}
\input{colordvi.tex}

\def\Mpl{M_{\rm pl}}
\newcommand{\FT}{{\cal F}_T}
\newcommand{\GT}{{\cal G}_T}
\newcommand{\e}{{\cal E}}
\newcommand{\p}{{\cal P}}

\usepackage{babel}

\makeatother

\usepackage{babel}
\begin{document}

\title{Effective gravitational couplings for cosmological perturbations
in the most general scalar-tensor theories with second-order field
equations}

\author{Antonio De Felice}

\affiliation{TPTP \& NEP, The Institute for Fundamental Study, Naresuan University,
Phitsanulok 65000, Thailand}

\affiliation{Thailand Center of Excellence in Physics, Ministry of Education,
Bangkok 10400, Thailand}

\affiliation{Department of Physics, Faculty of Science, Tokyo University of Science,
1-3, Kagurazaka, Shinjuku-ku, Tokyo 162-8601, Japan}

\author{Tsutomu Kobayashi}

\affiliation{Hakubi Center, Kyoto University, Kyoto 606-8302, Japan }

\affiliation{Department of Physics, Kyoto University, Kyoto 606-8502, Japan }

\author{Shinji Tsujikawa}

\affiliation{Department of Physics, Faculty of Science, Tokyo University of Science,
1-3, Kagurazaka, Shinjuku-ku, Tokyo 162-8601, Japan}

\date{\today}
\begin{abstract}
In the Horndeski's most general scalar-tensor theories the equations
of scalar density perturbations are derived in the presence of non-relativistic
matter minimally coupled to gravity. Under a quasi-static approximation
on sub-horizon scales we obtain the effective gravitational coupling
$G_{{\rm eff}}$ associated with the growth rate of matter perturbations
as well as the effective gravitational potential $\Phi_{{\rm eff}}$
relevant to the deviation of light rays. We then apply our formulas
to a number of modified gravitational models of dark energy--such
as those based on $f(R)$ theories, Brans-Dicke theories, kinetic
gravity braidings, covariant Galileons, and field derivative couplings
with the Einstein tensor. Our results are useful to test the large-distance
modification of gravity from the future high-precision observations
of large-scale structure, weak lensing, and cosmic microwave background.
\end{abstract}
\maketitle

\section{Introduction}

The late-time cosmic acceleration has been supported by several independent
observations--such as supernovae Ia \cite{SNIa}, Cosmic Microwave
Background (CMB) \cite{WMAP}, and baryon acoustic oscillations \cite{BAO}.
The simplest candidate for dark energy is the cosmological constant,
but the typical scale of the vacuum energy is vastly larger than the
observed energy scale of dark energy \cite{Weinberg}. Instead many
alternative models have been proposed to identify the origin of dark
energy \cite{review1,review2}.

A minimally coupled scalar field with a potential $V(\phi)$--quintessence
\cite{quin}--can account for the cosmic acceleration today, provided
that the potential is sufficiently flat with a small effective mass
$m_{\phi}\approx10^{-33}$\,eV. The k-essence \cite{kes}, where
the Lagrangian includes a nonlinear term of the field kinetic energy,
can be responsible for dark energy even in the absence of a field
potential. Quintessence and k-essence can be distinguished from the
cosmological constant in that their equations of state vary in time
while the latter does not.

There is another class of dark energy models based on the large-distance
modification of gravity--such as (i) $f(R)$ theories \cite{fR},
(ii) Brans-Dicke theories \cite{Brans}, (iii) Dvali--Gabadadze--Porrati
(DGP) braneworld \cite{DGP}, and (iv) Galileon gravity \cite{Nicolis}.
In the local region where the average density is much larger than
the cosmological one, these models need to recover the General Relativistic
behavior for consistency with solar-system experiments \cite{Soti}.
The models based on the theories (i) and (ii) can be made to be compatible
with local gravity constraints under the chameleon mechanism \cite{Khoury},
as long as the scalar degree of freedom has a large effective mass
in the region of high density \cite{fRlocal}. For the models based
on the theories (iii) and (iv) the nonlinear field self-interaction
can allow the recovery of the General Relativistic behavior in the
local region \cite{DGPVain} through the Vainshtein mechanism \cite{Vainshtein}.

For the dark energy models mentioned above the field equations of
motion are kept up to second order. This is desirable to avoid the
appearance of the Ostrogradski's instability \cite{Ost} associated
with the derivatives higher than the second order. In 1974 Horndeski
\cite{Horndeski} derived the most general single-field Lagrangian
for scalar-tensor theories with second-order equations of motion.
Recently this issue was revisited by Deffayet \textit{et al.} \cite{DGSZ}
in connection to a covariant Galileon field. The most general scalar-tensor
theories with the second-order equations can be expressed by the sum
of the Lagrangians (\ref{eachlag2})-(\ref{eachlag5}) below. In fact
one can show that this Lagrangian is equivalent to that derived by
Horndeski \cite{KYY} (see also Ref.~\cite{Char}).

The most general scalar-tensor theories not only include quintessence
and k-essence but also accommodate $f(R)$ theories, Brans-Dicke theories,
and Galileon gravity. Moreover, as shown in Ref.~\cite{KYY}, several
different choices of the functions $K$, $G_{i}$ ($i=3,4,5$) give
rise to the (modified) DGP model in 4 dimensions \cite{Turner}, the
field coupling with the Gauss-Bonnet term \cite{GBpapers}, the field-derivative
coupling with the Einstein tensor \cite{Germani,Linder}, and so on.

In this paper we shall derive the equations of linear density perturbations
for the most general scalar-tensor theories with non-relativistic
matter taken into account. In the presence of the terms ${\cal L}_{i}$
($i=3,4,5$) the effective gravitational coupling $G_{{\rm eff}}$
is subject to change compared to that in General Relativity. This
leads to the modified growth rate of matter density perturbations
$\delta_{m}$ as well as the modified evolution of the effective gravitational
potential $\Phi_{{\rm eff}}$ associated with the deviation of light
rays. Similar analysis has been carried out in specific scalar-tensor
theories \cite{Polarski,Ria,Verde,review2,Tsuji1,Koyama10}, $f(R)$
theories \cite{Zhang,Tsuji1}, kinetic gravity braidings with the
term ${\cal L}_{3}$ \cite{Chow,Koba10,Silva,Mukoh,Kimura}, and covariant
Galileon \cite{kase}. Our analysis in this paper covers those theories
as specific cases. Such general analysis will be useful to discriminate
between modified gravitational models from the observations of large-scale
structure, weak lensing, and CMB \cite{obsermo}.

This paper is organized as follows. In Sec.~\ref{backsec} we derive
the background equations of motion on the flat Friedmann-Lema\^{i}tre-Robertson-Walker
(FLRW) background for the action (\ref{action}) below. In Sec.~\ref{persec}
we obtain the full scalar perturbation equations of motion for the
metric (\ref{permet}). In Sec.~\ref{subhsec} the effective gravitational
coupling $G_{{\rm eff}}$ as well as the effective gravitational potential
$\Phi_{{\rm eff}}$ are derived under the quasi-static approximation
on sub-horizon scales. In Sec.~\ref{appsec} we apply our formulas
of sub-horizon perturbations to a number of modified gravitational
models of dark energy. Sec.~\ref{consec} is devoted to conclusions.

\section{The most general scalar-tensor theories and the background equations
of motion}

\label{backsec} The most general 4-dimensional scalar-tensor theories
keeping the field equations of motion at second order are described
by the Lagrangian \cite{DGSZ} 
\begin{equation}
{\cal L}=\sum_{i=2}^{5}{\cal L}_{i}\,,\label{Lagsum}
\end{equation}
 where 
\begin{eqnarray}
{\cal L}_{2} & = & K(\phi,X),\label{eachlag2}\\
{\cal L}_{3} & = & -G_{3}(\phi,X)\Box\phi,\\
{\cal L}_{4} & = & G_{4}(\phi,X)\, R+G_{4,X}\,[(\Box\phi)^{2}-(\nabla_{\mu}\nabla_{\nu}\phi)\,(\nabla^{\mu}\nabla^{\nu}\phi)]\,,\\
{\cal L}_{5} & = & G_{5}(\phi,X)\, G_{\mu\nu}\,(\nabla^{\mu}\nabla^{\nu}\phi)-\frac{1}{6}\, G_{5,X}\,[(\Box\phi)^{3}-3(\Box\phi)\,(\nabla_{\mu}\nabla_{\nu}\phi)\,(\nabla^{\mu}\nabla^{\nu}\phi)+2(\nabla^{\mu}\nabla_{\alpha}\phi)\,(\nabla^{\alpha}\nabla_{\beta}\phi)\,(\nabla^{\beta}\nabla_{\mu}\phi)]\,.\label{eachlag5}
\end{eqnarray}
 Here $K$ and $G_{i}$ ($i=3,4,5$) are functions in terms of a scalar
field $\phi$ and its kinetic energy $X=-\partial^{\mu}\phi\partial_{\mu}\phi/2$
with the partial derivatives $G_{i,X}\equiv\partial G_{i}/\partial X$,
$R$ is the Ricci scalar, and $G_{\mu\nu}$ is the Einstein tensor.
The above Lagrangian was first discovered by Horndeski in a different
form \cite{Horndeski}. In fact the Lagrangian (\ref{Lagsum}) is
equivalent to that derived by Horndeski \cite{KYY}.

We are interested in the late-time cosmology in which the field $\phi$
is responsible for dark energy. In addition we take into account a
barotropic perfect fluid with the equation of state $w=P_{m}/\rho_{m}$,
where $P_{m}$ is the pressure and $\rho_{m}$ is the energy density
respectively. In the following we focus on non-relativistic matter
($w=0$) minimally coupled to the field $\phi$.%
\footnote{If matter is non-minimally coupled to $\phi$ through the coupling
to the metric $\tilde{g}_{\mu\nu}=A(\phi)g_{\mu\nu}$ rather than
$g_{\mu\nu}$, one rewrites the action~(\ref{action}) in terms of
$\phi$ and $\tilde{g}_{\mu\nu}$ rather than $\phi$ and $g_{\mu\nu}$.
Then, the Lagrangian is still of the form~(\ref{eachlag2})--(\ref{eachlag5}),
because the change of the variable $g_{\mu\nu}\to\tilde{g}_{\mu\nu}=A(\phi)g_{\mu\nu}$
does not generate higher derivative terms in the field equations.
For this reason, it is sufficient to consider matter minimally coupled
to $\phi$. %
} The total action we are going to study is then given by 
\begin{equation}
S=\int d^{4}x\sqrt{-g}\left({\cal L}+{\cal L}_{m}\right)\,,\label{action}
\end{equation}
 where $g$ is a determinant of the metric $g_{\mu\nu}$, and ${\cal L}_{m}$
is the Lagrangian of non-relativistic matter.

Let us consider a flat FLRW background with the metric $ds^{2}=-N^{2}(t)dt^{2}+a^{2}(t)d{\bm{x}}^{2}$.
Variations with respect to the lapse $N(t)$ and the scale factor
$a(t)$ give rise to the following equations of motion respectively
\begin{eqnarray}
 &  & \e\equiv\sum_{i=2}^{5}\e_{i}=-\rho_{m}\,,\label{be1}\\
 &  & \p\equiv\sum_{i=2}^{5}\p_{i}=0\,,\label{be2}
\end{eqnarray}
 where 
\begin{eqnarray}
\e_{2} & \equiv & 2XK_{,X}-K,\\
\e_{3} & \equiv & 6X\dot{\phi}HG_{3,X}-2XG_{3,\phi},\\
\e_{4} & \equiv & -6H^{2}G_{4}+24H^{2}X(G_{4,X}+XG_{4,XX})-12HX\dot{\phi}G_{4,\phi X}-6H\dot{\phi}G_{4,\phi}\,,\\
\e_{5} & \equiv & 2H^{3}X\dot{\phi}\left(5G_{5,X}+2XG_{5,XX}\right)-6H^{2}X\left(3G_{5,\phi}+2XG_{5,\phi X}\right)\,,
\end{eqnarray}
 and 
\begin{eqnarray}
{\cal P}_{2} & \equiv & K,\\
{\cal P}_{3} & \equiv & -2X\left(G_{3,\phi}+\ddot{\phi}\, G_{3,X}\right),\\
{\cal P}_{4} & \equiv & 2\left(3H^{2}+2\dot{H}\right)G_{4}-12H^{2}XG_{4,X}-4H\dot{X}G_{4,X}-8\dot{H}XG_{4,X}-8HX\dot{X}G_{4,XX}+2\left(\ddot{\phi}+2H\dot{\phi}\right)G_{4,\phi}\nonumber \\
 &  & +4XG_{4,\phi\phi}+4X\left(\ddot{\phi}-2H\dot{\phi}\right)G_{4,\phi X},\\
{\cal P}_{5} & \equiv & -2X\left(2H^{3}\dot{\phi}+2H\dot{H}\dot{\phi}+3H^{2}\ddot{\phi}\right)G_{5,X}-4H^{2}X^{2}\ddot{\phi}\, G_{5,XX}+4HX\left(\dot{X}-HX\right)G_{5,\phi X}\nonumber \\
 &  & +2\left[2\left(\dot{H}X+H\dot{X}\right)+3H^{2}X\right]G_{5,\phi}+4HX\dot{\phi}\, G_{5,\phi\phi}.
\end{eqnarray}
 Here a dot represents a derivative with respect to $t$ and $H\equiv\dot{a}/a$
is the Hubble parameter. Varying the action (\ref{action}) with respect
to $\phi(t)$, it follows that 
\begin{equation}
\frac{1}{a^{3}}\frac{d}{dt}\left(a^{3}J\right)=P_{\phi}\,,\label{fieldeq}
\end{equation}
 where 
\begin{eqnarray}
J & \equiv & \dot{\phi}K_{,X}+6HXG_{3,X}-2\dot{\phi}G_{3,\phi}+6H^{2}\dot{\phi}\left(G_{4,X}+2XG_{4,XX}\right)-12HXG_{4,\phi X}\nonumber \\
 &  & +2H^{3}X\left(3G_{5,X}+2XG_{5,XX}\right)-6H^{2}\dot{\phi}\left(G_{5,\phi}+XG_{5,\phi X}\right)\,,\\
P_{\phi} & \equiv & K_{,\phi}-2X\left(G_{3,\phi\phi}+\ddot{\phi}\, G_{3,\phi X}\right)+6\left(2H^{2}+\dot{H}\right)G_{4,\phi}+6H\left(\dot{X}+2HX\right)G_{4,\phi X}\nonumber \\
 &  & -6H^{2}XG_{5,\phi\phi}+2H^{3}X\dot{\phi}\, G_{5,\phi X}\,.
\end{eqnarray}
 Non-relativistic matter obeys the continuity equation 
\begin{equation}
\dot{\rho}_{m}+3H\rho_{m}=0\,.\label{rhoeq}
\end{equation}
 Equations (\ref{be1}), (\ref{be2}), (\ref{fieldeq}), and (\ref{rhoeq})
are not independent because of the Bianchi identities. In fact the
field equation (\ref{fieldeq}) can be derived by using Eqs.~(\ref{be1}),
(\ref{be2}), and (\ref{rhoeq}).

\section{Linear perturbation equations}

\label{persec} In this section we derive the linear perturbation
equations for the theories give by the action (\ref{action}). Let
us consider the following perturbed metric about the flat FLRW background
\cite{Bardeen} 
\begin{equation}
ds^{2}=-(1+2\Psi)\, dt^{2}-2\partial_{i}\chi dtdx^{i}+a^{2}(t)(1+2\Phi)\delta_{ij}dx^{i}dx^{j}\,,\label{permet}
\end{equation}
where $\Psi,\Phi$, and $\chi$ are the scalar metric perturbations.
In this expression we have chosen a spatial gauge such that $g_{ij}$
is diagonal, which fixes the spatial part of a vector associated with
a scalar gauge transformation. The temporal part of the gauge-transformation
vector is not fixed for the moment.

We perturb the scalar field as $\phi(t)+\delta\phi(t,{\bm{x}})$,
and the matter fields as well, in terms of the matter density perturbation
$\delta\rho_{m}$, and the scalar part of the fluid velocity $v$.
We define the density contrast of matter as $\delta\equiv\delta\rho_{m}/\rho_{m}$.
In order to write the perturbation equations in a compact form, we
introduce the following quantities \cite{KYY}, 
\begin{eqnarray}
\FT & \equiv & 2\left[G_{4}-X\left(\ddot{\phi}\, G_{5,X}+G_{5,\phi}\right)\right]\,,\\
\GT & \equiv & 2\left[G_{4}-2XG_{4,X}-X\left(H\dot{\phi}\, G_{5,X}-G_{5,\phi}\right)\right]\,,
\end{eqnarray}
 and 
\begin{eqnarray}
\Theta & \equiv & -\frac{1}{6}\frac{\partial{\cal E}}{\partial H}\nonumber \\
 & = & -\dot{\phi}XG_{3,X}+2HG_{4}-8HXG_{4,X}-8HX^{2}G_{4,XX}+\dot{\phi}G_{4,\phi}+2X\dot{\phi}\, G_{4,\phi X}\nonumber \\
 &  & -H^{2}\dot{\phi}\left(5XG_{5,X}+2X^{2}G_{5,XX}\right)+2HX\left(3G_{5,\phi}+2XG_{5,\phi X}\right)\,,\\
\Sigma & \equiv & X\frac{\partial{\cal E}}{\partial X}+\frac{1}{2}H\frac{\partial{\cal E}}{\partial H}\nonumber \\
 & = & XK_{,X}+2X^{2}K_{,XX}+12H\dot{\phi}XG_{3,X}+6H\dot{\phi}X^{2}G_{3,XX}-2XG_{3,\phi}-2X^{2}G_{3,\phi X}\nonumber \\
 &  & -6H^{2}G_{4}+6\left[H^{2}\left(7XG_{4,X}+16X^{2}G_{4,XX}+4X^{3}G_{4,XXX}\right)-H\dot{\phi}\left(G_{4,\phi}+5XG_{4,\phi X}+2X^{2}G_{4,\phi XX}\right)\right]\nonumber \\
 &  & +30H^{3}\dot{\phi}XG_{5,X}+26H^{3}\dot{\phi}X^{2}G_{5,XX}+4H^{3}\dot{\phi}X^{3}G_{5,XXX}-6H^{2}X\bigl(6G_{5,\phi}+9XG_{5,\phi X}+2X^{2}G_{5,\phi XX}\bigr)\,,
\end{eqnarray}
 where we used the relation $X\partial_{X}\dot{\phi}=\dot{\phi}/2$.
The functions $\FT$ and $\GT$ appear in the quadratic action for
the cosmological tensor perturbations \cite{KYY,Gao,infnon}. In order
to avoid ghost and Laplacian instabilities in the tensor sector we
require the conditions $\FT>0$ and $\GT>0$.

We expand the action (\ref{action}) up to second order in perturbations
and vary the second-order action with respect to each perturbed variable
such as $\Psi$. Following the procedure explained in Ref.~\cite{kase},
the perturbation equations in Fourier space are given by 
\begin{eqnarray}
E_{\Psi} & \equiv & A_{1}\dot{\Phi}+A_{2}\dot{\delta\phi}-\rho_{m}\dot{v}+A_{3}\frac{k^{2}}{a^{2}}\Phi+A_{4}\Psi+A_{5}\frac{k^{2}}{a^{2}}\chi+\left(A_{6}\frac{k^{2}}{a^{2}}-\mu\right)\delta\phi-\rho_{m}\delta=0\,,\label{eq:Psi}\\
E_{\Phi} & \equiv & B_{1}\ddot{\Phi}+B_{2}\ddot{\delta\phi}+B_{3}\dot{\Phi}+B_{4}\dot{\delta\phi}+B_{5}\dot{\Psi}+B_{6}\frac{k^{2}}{a^{2}}\Phi+\left(B_{7}\frac{k^{2}}{a^{2}}+3\nu\right)\delta\phi\nonumber \\
 &  & {}+\left(B_{8}\frac{k^{2}}{a^{2}}+B_{9}\right)\Psi+B_{10}\frac{k^{2}}{a^{2}}\dot{\chi}+B_{11}\frac{k^{2}}{a^{2}}\chi+3\rho_{m}\dot{v}=0\,,\label{eq:Phi}\\
E_{\chi} & \equiv & C_{1}\dot{\Phi}+C_{2}\dot{\delta\phi}+C_{3}\Psi+C_{4}\delta\phi+\rho_{m}v=0\,,\\
E_{\delta\phi} & \equiv & D_{1}\ddot{\Phi}+D_{2}\ddot{\delta\phi}+D_{3}\dot{\Phi}+D_{4}\dot{\delta\phi}+D_{5}\dot{\Psi}+D_{6}\frac{k^{2}}{a^{2}}\dot{\chi}\nonumber \\
 &  & {}+\left(D_{7}\frac{k^{2}}{a^{2}}+D_{8}\right)\Phi+\left(D_{9}\frac{k^{2}}{a^{2}}-M^{2}\right)\delta\phi+\left(D_{10}\frac{k^{2}}{a^{2}}+D_{11}\right)\Psi+D_{12}\frac{k^{2}}{a^{2}}\chi=0\,,\label{eq:dPi}\\
E_{v} & \equiv & \dot{v}-\Psi=0\,,\label{eq:mat1}\\
E_{\delta} & \equiv & \dot{\delta}+3\dot{\Phi}+\frac{k^{2}}{a^{2}}v-\frac{k^{2}}{a^{2}}\chi=0\,,\label{eq:mat2}
\end{eqnarray}
 where $k$ is a comoving wavenumber, and 
\begin{eqnarray}
 &  & A_{1}=6\Theta,\qquad A_{2}=-2(\Sigma+3H\Theta)/\dot{\phi},\qquad A_{3}=2\GT,\qquad A_{4}=2\Sigma+\rho_{m},\qquad A_{5}=-2\Theta,\nonumber \\
 &  & A_{6}=2(\Theta-H\GT)/\dot{\phi},\qquad\mu={\cal E}_{,\phi}\,,\label{Ai}
\end{eqnarray}
 
\begin{eqnarray}
 &  & B_{1}=6\GT,\qquad B_{2}=6(\Theta-H\GT)/\dot{\phi},\qquad B_{3}=6(\dot{\GT}+3H\GT),\nonumber \\
 &  & B_{4}=3\left[\left(4H\ddot{\phi}-4\dot{H}\dot{\phi}-6H^{2}\dot{\phi}\right)\GT-2H\dot{\phi}\,\dot{\GT}-\left(4\ddot{\phi}-6H\dot{\phi}\right)\Theta+2\dot{\phi}\dot{\Theta}-\rho_{m}\dot{\phi}\right]/\dot{\phi}^{2},\qquad B_{5}=-6\Theta,\nonumber \\
 &  & B_{6}=2\FT,\qquad B_{7}=2\left[\dot{\GT}+H\left(\GT-\FT\right)\right]/\dot{\phi},\qquad B_{8}=2\GT,\qquad B_{9}=-6(\dot{\Theta}+3H\Theta),\nonumber \\
 &  & B_{10}=-2\GT,\qquad B_{11}=-2(\dot{\GT}+H\GT),\qquad\nu={\cal P}_{,\phi}\,,\label{Bi}
\end{eqnarray}
 
\begin{equation}
C_{1}=2\GT,\qquad C_{2}=2(\Theta-H\GT)/\dot{\phi},\qquad C_{3}=-2\Theta,\qquad C_{4}=\left[2(H\ddot{\phi}-\dot{H}\dot{\phi})\GT-2\ddot{\phi}\,\Theta-\rho_{m}\dot{\phi}\right]/\dot{\phi}^{2},\label{Ci}
\end{equation}
\begin{eqnarray}
 &  & D_{1}=6(\Theta-H\GT)/\dot{\phi},\qquad D_{2}=2(3H^{2}\GT-6H\Theta-\Sigma)/\dot{\phi}^{2},\nonumber \\
 &  & D_{3}=-3\left[2H(\dot{\GT}+3H\GT)-2(\dot{\Theta}+3H\Theta)-\rho_{m}\right]/\dot{\phi},\nonumber \\
 &  & D_{4}=2\left[3H\{(3H^{2}+2\dot{H})\dot{\phi}-2H\ddot{\phi}\}\GT+3H^{2}\dot{\phi}\dot{\GT}+6\{2H\ddot{\phi}-(3H^{2}+\dot{H})\dot{\phi}\}\Theta-6H\dot{\phi}\dot{\Theta}+(2\ddot{\phi}-3H\dot{\phi})\Sigma-\dot{\phi}\dot{\Sigma}\right]/\dot{\phi}^{3},\nonumber \\
 &  & D_{5}=2(\Sigma+3H\Theta)/\dot{\phi},\qquad D_{6}=-2(\Theta-H\GT)/\dot{\phi},\qquad D_{7}=2\left[\dot{\GT}+H\left(\GT-\FT\right)\right]/\dot{\phi},\nonumber \\
 &  & D_{8}=3\left[6(\dot{H}\dot{\phi}-H\ddot{\phi})\Theta-2\ddot{\phi}\Sigma+3H\rho_{m}\dot{\phi}-\mu\dot{\phi}^{2}\right]/\dot{\phi}^{2},\nonumber \\
 &  & D_{9}=\left[2H^{2}\FT-4H(\dot{\GT}+H\GT)+2(\dot{\Theta}+H\Theta)+\rho_{m}\right]/\dot{\phi}^{2},\nonumber \\
 &  & D_{10}=2(\Theta-H\GT)/\dot{\phi},\qquad D_{11}=\left[6\{(3H^{2}+\dot{H})\dot{\phi}-H\ddot{\phi}\}\Theta+6H\dot{\phi}\dot{\Theta}+2(3H\dot{\phi}-\ddot{\phi})\Sigma+2\dot{\phi}\dot{\Sigma}-\mu\dot{\phi}^{2}\right]/\dot{\phi}^{2},\nonumber \\
 &  & D_{12}=\left[2H(\dot{\GT}+H\GT)-2(\dot{\Theta}+H\Theta)-\rho_{m}\right]/\dot{\phi}\,,\nonumber \\
 &  & M^{2}=\left[\dot{\mu}+3H(\mu+\nu)\right]/\dot{\phi}\nonumber \\
 &  & ~~~~~\,=-K_{,{\phi\phi}}+(\ddot{\phi}+3H\dot{\phi})K_{,\phi X}+2XK_{,\phi\phi X}+2X\ddot{\phi}K_{,\phi XX}\nonumber \\
 &  & ~~~~~~~~~{}+[6\, H(G_{{3,\phi{\it XX}}}X+G_{{3,\phi X}})\dot{\phi}-2\, G_{{3,\phi\phi X}}X-2\, G_{{3,\phi\phi}}]\ddot{\phi}+6\, H\left(G_{{3,\phi\phi X}}X-G_{{3,\phi\phi}}\right)\dot{\phi}\nonumber \\
 &  & ~~~~~~~~~{}+6\, G_{{3,\phi X}}X\dot{H}+2(9\,{H}^{2}G_{{3,\phi X}}-G_{{3,\phi\phi\phi}})X\nonumber \\
 &  & ~~~~~~~~~{}+[6\,{H}^{2}(4\, G_{{4,\phi{\it XXX}}}{X}^{2}+8\, G_{{4,\phi{\it XX}}}X+G_{{4,\phi X}})-6\, H(2\, G_{{4,\phi\phi{\it XX}}}X+3\, G_{{4,\phi\phi X}})\dot{\phi}]\ddot{\phi}\nonumber \\
 &  & ~~~~~~~~~{}+[12\, H(G_{{4,\phi X}}+2\, G_{{4,\phi{\it XX}}}X)\dot{H}+6\, H(6\,{H}^{2}G_{{4,\phi{\it XX}}}X-2\, G_{{4,\phi\phi\phi X}}X+3\,{H}^{2}G_{{4,\phi X}})]\dot{\phi}\nonumber \\
 &  & ~~~~~~~~~{}+12\,{H}^{2}\left(2\, G_{{4,\phi\phi{\it XX}}}{X}^{2}-3\, G_{{4,\phi\phi X}}X-G_{{4,\phi\phi}}\right)-6\left(2\, G_{{4,\phi\phi X}}X+G_{{4,\phi\phi}}\right)\dot{H}\nonumber \\
 &  & ~~~~~~~~~{}+[2\,{H}^{3}(2\, G_{{5,\phi{\it XXX}}}{X}^{2}+7\, G_{{5,\phi{\it XX}}}X+3\, G_{{5,\phi X}})\dot{\phi}-6\,{H}^{2}(5\, G_{{5,\phi\phi X}}X+G_{{5,\phi\phi}}+2\, G_{{5,\phi\phi{\it XX}}}{X}^{2})]\ddot{\phi}\nonumber \\
 &  & ~~~~~~~~~{}+[2H^{3}(2\, G_{{5,\phi\phi{\it XX}}}{X}^{2}-9\, G_{{5,\phi\phi}}-7\, G_{{5,\phi\phi X}}X)-12\, H(G_{{5,\phi\phi X}}X+G_{{5,\phi\phi}})\dot{H}]\dot{\phi}\nonumber \\
 &  & ~~~~~~~~~{}+6\,{H}^{2}X\left(3\, G_{{5,\phi X}}+2\, G_{{5,\phi{\it XX}}}X\right)\dot{H}+6\,{H}^{2}X\left(3\,{H}^{2}G_{{5,\phi X}}-G_{{5,\phi\phi\phi}}+2\,{H}^{2}G_{{5,\phi{\it XX}}}X-2\, G_{{5,\phi\phi\phi X}}X\right).\label{masss}
\end{eqnarray}
In deriving the above we used the background equations. The expression
of the coefficients of the equations (\ref{eq:Psi})-(\ref{eq:dPi}),
written in terms of the variables $\Theta$, $\Sigma$, etc., becomes
compact, though many of the coefficients include terms $\dot{\phi}^{n}$
($n>0$) in the denominators. However, there are no divergences at
$\dot{\phi}=0$. In fact, whenever the term $\dot{\phi}^{n}$ appears
in the denominator, the numerator of the same coefficient compensates
with the term $\dot{\phi}^{n}$. For instance, this property can be
seen in the expression of $M^{2}$. From the expressions
of the coefficients given above, it is clear that not all these coefficients
are independent. For example, later on, we will find it convenient
to use the following relations $A_{3}=B_{8}$, $D_{7}=B_{7}$, and
$D_{10}=A_{6}$.

The mass of the field $\delta\phi$ is related with the term $M^{2}$
defined in Eq.~(\ref{masss}). In fact, for a canonical scalar field described
by the Lagrangian $K=X-V(\phi)$ with $G_{i}=0$ ($i=3,4,5$), we
have that $M^{2}=-K_{,\phi \phi}=V_{,\phi\phi}$. 
In viable dark energy models based on $f(R)$ gravity and Brans-Dicke theory with 
a field potential, the term $-K_{,\phi \phi}$ is the dominant contribution 
to $M^2$ \cite{TsujifR,Tsujistensor}.
The term $-K_{,\phi \phi}$ comes from the time-derivative of $\mu$, 
such that the contribution from the term 
$3H(\mu+\nu)$ is usually unimportant relative to $\dot{\mu}$.

The equations of motion (\ref{eq:Psi})-(\ref{eq:mat2}) are not independent. 
In fact, we find the identity  
\begin{equation}
\dot{E}_{\Psi}+3HE_{\Psi}-HE_{\Phi}-\frac{k^{2}}{a^{2}}E_{\chi}-\dot{\phi}E_{\delta\phi}+\rho_{m}(\dot{E}_{v}+3HE_{v})+\rho_{m}E_{\delta}=0\,.\label{eq:bian1}
\end{equation}
 This relation can be used in two ways: 1) to check the consistency
of the equations themselves; 2) to get some equations of motion, which
would be missing when some gauge is used from the beginning. For example,
in the Newtonian gauge ($\chi=0$), the equation $\left.E_{\chi}\right|_{\chi=0}=0$,
cannot be derived directly. However, it is still possible to obtain
it by using Eq.~(\ref{eq:bian1}).

We note that the following combination of the perturbation equations
is useful: 
\begin{equation}
\frac{k^{2}}{a^{2}}\tilde{E}_{\gamma}\equiv3\left(\dot{E}_{\chi}+3HE_{\chi}\right)-E_{\Phi}=0\,,\label{eq:Gam}
\end{equation}
 which is written explicitly as 
\begin{equation}
\tilde{E}_{\gamma}=B_{6}\Phi+B_{7}\delta\phi+B_{8}\Psi+B_{10}\dot{\chi}+B_{11}\chi=0\,.\label{eq:gamma}
\end{equation}
 This equation corresponds to the traceless part of the gravitational
field equations.

In order to study the evolution of matter perturbations we introduce
the gauge-invariant density contrast 
\begin{equation}
\delta_{m}\equiv\delta+3Hv\,.
\end{equation}
 {}From Eqs.~(\ref{eq:mat1}) and (\ref{eq:mat2}) it follows that
\begin{equation}
\ddot{\delta}_{m}+2H\dot{\delta}_{m}+\frac{k^{2}}{a^{2}}(\Psi-\dot{\chi})=3\left(\ddot{I}+2H\dot{I}\right)\,,\label{deltam}
\end{equation}
 where $I\equiv Hv-\Phi$.

\section{Effective gravitational couplings under the quasi-static approximation
on sub-horizon scales}
\label{subhsec} 

When we discuss the evolution of matter perturbations
relevant to large-scale structure and weak lensing, we are primarily
interested in the modes deep inside the Hubble radius ($k^{2}/a^{2}\gg H^{2}$).
We shall use the quasi-static approximation on sub-horizon scales,\footnote{Strictly
speaking, the typical scale here should be given by the sound horizon rather than the Hubble horizon
because the propagation speed of the scalar mode, $c_s$, 
differs from unity in general~\cite{newpaper, KYY}.
One needs to be careful for the use of the quasi-static approximation
in models with $c_s\ll 1$, as the range of the validity of the approximation may be quite limited.}
under which the dominant contributions in the perturbation equations
correspond to those including $k^{2}/a^{2}$ and $\delta$ \cite{Star98,Polarski,review2}.
There are two different classes of dark energy models, depending 
on the mass $M$ of a scalar degree of freedom.

The first one corresponds to the case in which the mass 
$M$ becomes large in the early cosmological epoch. 
The viable dark energy models constructed
in the framework of $f(R)$ gravity \cite{fRviable,fRStar,TsujifR}
and Brans-Dicke theories \cite{Tsujistensor} belong to this class.
Since the effect of the field mass cannot be neglected in such cases, 
we need to take into account the term $M^{2}$ to discuss
the evolution of perturbations.
This induces the oscillation of the field perturbation $\delta \phi$, 
but as long as this oscillating mode
is initially suppressed relative to the matter-induced mode, 
the quasi-static approximation can reproduce numerically 
integrated solutions with high accuracy \cite{TsujifR,fRosc}.
Under the quasi-static approximation we neglect the time-derivatives 
of $\delta \phi$, which corresponds to 
the approximation under which the oscillating mode is unimportant 
relative to the matter-induced mode.

Another class corresponds the case in which the field does not 
have a massive potential, e.g., Galileon gravity.
In such cases the numerical simulations in Refs.~\cite{Koba10,kase} 
also show that the quasi-static approximation is sufficiently accurate 
for the modes deep inside the Hubble radius.

We expect that the quasi-static approximation on sub-horizon scales 
should be trustable for our general theories as well, provided that 
the matter-induced mode dominates over the oscillating mode.

Let us choose the Newtonian gauge in which $\chi=0$. Under the quasi-static
approximation on sub-horizon scales we find that Eqs.~(\ref{eq:gamma}),
(\ref{eq:dPi}), and (\ref{eq:Psi}) can be rewritten as 
\begin{eqnarray}
B_{6}\Phi+B_{7}\delta\phi+B_{8}\Psi & = & 0\,,\label{eq:axg1}\\
B_{7}\frac{k^{2}}{a^{2}}\Phi+\left(D_{9}\frac{k^{2}}{a^{2}}
-M^{2}\right)\delta\phi+A_{6}\frac{k^{2}}{a^{2}}\Psi & \simeq & 0\,,\label{eq:axg2}\\
B_{8}\frac{k^{2}}{a^{2}}\Phi+A_{6}\frac{k^{2}}{a^{2}}
\delta\phi-\rho_{m}\delta & \simeq & 0\,,\label{eq:axg3}
\end{eqnarray}
 where we used the relations $A_{3}=B_{8}$,
$D_{7}=B_{7}$, and $D_{10}=A_{6}$ already mentioned
in Sec.~\ref{persec}. 
The above three equations correspond
to the traceless part of the gravitational field equations, the scalar-field
equation of motion, and the $(00)$-component of the gravitational
field equations, respectively. 
Note that Eq.~(\ref{eq:axg1}) also follows from Eq.~(\ref{eq:Phi})
in the same approximation scheme.

We can solve Eqs.~(\ref{eq:axg1})
and~(\ref{eq:axg2}) for $\Phi$ and $\delta\phi$ in terms of $\Psi$,
and then substitute these expressions into Eq.~(\ref{eq:axg3}).
This gives the following Poisson equation 
\begin{equation}
\frac{k^{2}}{a^{2}}\Psi\simeq-4\pi G_{{\rm eff}}\rho_{m}\delta\,.\label{Psiap}
\end{equation}
 Here the effective gravitational coupling $G_{{\rm eff}}$ is given
by 
\begin{eqnarray}
G_{{\rm eff}} & = & {\frac{2\Mpl^{2}[(B_{6}D_{9}-B_{7}^{2})\,{(k/a)}^{2}-B_{6}M^{2}]}{(A_{6}^{2}B_{6}+B_{8}^{2}D_{9}-2A_{6}B_{7}B_{8})\,{(k/a)}^{2}-B_{8}^{2}M^{2}}}G\label{Geff1}\\
 & = & \frac{\Mpl^{2}\left\{ \left(\dot{\Theta}+H\Theta\right){\cal F}_{S}+\left(\dot{\GT}-\dot{\Theta}\GT/\Theta\right)^{2}+\FT\left[XM^{2}a^{2}/k^{2}+({\cal E}+{\cal P})/2\right]\right\} }{\Theta^{2}{\cal F}_{S}+\GT^{2}\left[XM^{2}a^{2}/k^{2}+({\cal E}+{\cal P})/2\right]}G\,,\label{eq:Geff}
\end{eqnarray}
 where $G$ is the bare gravitational constant related with the reduced
Planck mass $M_{{\rm pl}}$ via the relation $8\pi G=M_{{\rm pl}}^{-2}$,
and 
\begin{equation}
{\cal F}_{S}\equiv\frac{1}{a}\frac{d}{dt}\left(\frac{a}{\Theta}\GT^{2}\right)-\FT\,.
\end{equation}
 In order to avoid the Laplacian instability of scalar perturbations
we require that ${\cal F}_{S}>0$ \cite{KYY,Gao,infnon}. While $G_{{\rm eff}}$
is written in a compact expression in Eq.~(\ref{eq:Geff}), it is
often convenient to use the form (\ref{Geff1}) for a given Lagrangian.
In Appendix A we present the explicit forms of the coefficients $A_{6}$,
$B_{6}$, $B_{7}$, $B_{8}$, and $D_{9}$, which is useful for the
computation of Eq.~(\ref{Geff1}).

Under the quasi-static approximation on sub-horizon scales Eq.~(\ref{deltam})
gives 
\begin{equation}
\ddot{\delta}_{m}+2H\dot{\delta}_{m}+\frac{k^{2}}{a^{2}}\Psi\simeq0\,,
\end{equation}
On using Eq.~(\ref{Psiap}) and $\delta_{m}\simeq\delta$ (which
are valid for $k^{2}/a^{2}\gg H^{2}$), it follows that 
\begin{equation}
\ddot{\delta}_{m}+2H\dot{\delta}_{m}-4\pi G_{{\rm eff}}\rho_{m}\delta_{m}\simeq0\,.
\end{equation}
 This can be written as 
\begin{equation}
\delta_{m}''+\left(2+\frac{H'}{H}\right)\delta_{m}'-\frac{3}{2}\frac{G_{{\rm eff}}}{G}\Omega_{m}\delta_{m}\simeq0\,,\label{delmeq}
\end{equation}
 where $\Omega_{m}\equiv\rho_{m}/(3\Mpl^{2}H^{2})$, and a prime represents
a derivative with respect to $N=\ln a$.

We define the anisotropic parameter $\eta$ to characterize the difference
between the two gravitational potentials: 
\begin{equation}
\eta\equiv-\Phi/\Psi\,.\label{etadef}
\end{equation}
Under the quasi-static approximation on sub-horizon scales this reduces to 
\begin{eqnarray}
\eta & \simeq & \frac{(B_{8}D_{9}-A_{6}B_{7})(k/a)^{2}-B_{8}M^{2}}{(B_{6}D_{9}-B_{7}^{2})(k/a)^{2}-B_{6}M^{2}}\label{eta1}\\
 & = & \frac{\GT\dot{\Theta}+(H\FT-\dot{\GT})\Theta-H\GT(\dot{\GT}+H\GT)-\GT\left[({\cal E}+{\cal P})/2+XM^{2}a^{2}/k^{2}\right]}{\FT(\dot{\Theta}+H\Theta)-(\dot{\GT}+H\GT)^{2}-\FT\left[({\cal E}+{\cal P})/2+XM^{2}a^{2}/k^{2}\right]}\,.\label{eq:eta}
\end{eqnarray}
 We also introduce the effective gravitational potential 
\begin{equation}
\Phi_{{\rm eff}}\equiv(\Psi-\Phi)/2\,,
\end{equation}
 which is associated with the deviation of the light rays in CMB and
weak lensing observations \cite{Schmidt}. {}From Eqs.~(\ref{Psiap})
and (\ref{etadef}) we have 
\begin{equation}
\Phi_{{\rm eff}}\simeq-4\pi G_{{\rm eff}}\frac{1+\eta}{2}\left(\frac{a}{k}\right)^{2}\rho_{m}\delta\simeq-\frac{3}{2}\frac{G_{{\rm eff}}}{G}\frac{1+\eta}{2}\left(\frac{aH}{k}\right)^{2}\Omega_{m}\delta_{m}\,.\label{Phieff}
\end{equation}
 Let us consider k-essence in the framework of General Relativity
(GR), which corresponds to the Lagrangian ${\cal L}=K(\phi,X)+(\Mpl^{2}/2)R$
{[}i.e. $G_{4}=\Mpl^{2}/2$, $G_{3}=0=G_{5}${]}. In this case one
has $\FT=\GT=\Mpl^{2}$, $\Theta=H\Mpl^{2}$, and ${\cal F}_{S}=-\Mpl^{2}\dot{H}/H^{2}$,
which gives $G_{{\rm eff}}=G$ and $\eta=1$ from Eqs.~(\ref{eq:Geff})
and (\ref{eq:eta}). During the matter-dominated epoch ($H'/H\simeq-3/2$
and $\Omega_{m}\simeq1$) there is a growing mode solution $\delta_{m}\propto a$
to Eq.~(\ref{delmeq}). For this solution $\Phi_{{\rm eff}}=$\,constant
from Eq.~(\ref{Phieff}). In modified gravitational theories $G_{{\rm eff}}$
and $\eta$ are in general different from $G$ and $1$ respectively,
so that the evolution of $\delta_{m}$ and $\Phi_{{\rm eff}}$ is
subject to change compared to GR.

\section{Application to specific theories}

\label{appsec} In this section we apply our formulas of $G_{{\rm eff}}$
and $\eta$ derived in Sec.~\ref{subhsec} to a number of modified
gravitational theories.

\subsection{$f(R)$ theories}

The Lagrangian of $f(R)$ theories corresponds to ${\cal L}=(\Mpl^{2}/2)\, f(R)$,
where $f$ is an arbitrary function in terms of the Ricci scalar $R$.
This is equivalent to the Lagrangian (\ref{Lagsum}) by choosing the
following functions \cite{Ohanlon} 
\begin{equation}
K=-\frac{\Mpl^{2}}{2}\,(Rf_{,R}-f)\,,\qquad G_{3}=0=G_{5}\,,\qquad G_{4}=\frac{1}{2}\Mpl\phi\,,\qquad\phi=\Mpl f_{,R}\,,
\end{equation}
where $\phi$ is a scalar degree of freedom having the dimension of
mass. Since $G_{4}=\phi\Mpl/2$, $B_{7}=2A_{6}=2\Mpl$, $B_{6}=B_{8}=2\Mpl\phi$,
and $D_{9}=0$ in this case, Eqs.~(\ref{Geff1}) and (\ref{eta1})
read 
\begin{equation}
G_{{\rm eff}}=\frac{\Mpl}{\phi}\frac{4+2(\phi/\Mpl)(Ma/k)^{2}}{3+2(\phi/\Mpl)(Ma/k)^{2}}G\,,\quad\quad\quad\eta=\frac{1+(\phi/\Mpl)(Ma/k)^{2}}{2+(\phi/\Mpl)(Ma/k)^{2}}\,.\label{GefffR}
\end{equation}
{}From Eq.~(\ref{masss}) the mass squared of the scalar degree
of freedom is given by 
\begin{equation}
M^{2}=-K_{,\phi\phi}=\frac{1}{2f_{,RR}}\,.
\end{equation}
Substituting this relation and $\phi=\Mpl f_{,R}$ into Eq.~(\ref{GefffR}),
it follows that 
\begin{equation}
G_{{\rm eff}}=\frac{G}{f_{,R}}\,\frac{1+4(f_{,RR}/f_{,R})(k/a)^{2}}{1+3(f_{,RR}/f_{,R})(k/a)^{2}}\,,\quad\quad\quad\eta=\frac{1+2(f_{,RR}/f_{,R})(k/a)^{2}}{1+4(f_{,RR}/f_{,R})(k/a)^{2}}\,,\label{geffeta}
\end{equation}
 which agree with those derived in Ref.~\cite{Tsuji1}.

The viable dark energy models based on $f(R)$ theories were constructed
to have a large mass $M$ in the deep matter-dominated epoch \cite{fRviable,fRStar,TsujifR},
i.e. $f_{,RR}\gg1$ and $f_{,R}\simeq1$ for $R\gg H_{0}^{2}$, where
$H_{0}$ is the Hubble parameter today. In the regime $(f_{,RR}/f_{,R})(k/a)^{2}\ll1$
(or $M^{2}f_{,R}\gg k^{2}/a^{2}$) one has $G_{{\rm eff}}\simeq G$
and $\eta\simeq1$, so that the evolution of density perturbations
is similar to that in GR. At late times the mass term $M$ gets smaller
with the growth of $f_{,RR}$. Since $G_{{\rm eff}}\simeq4G/(3f_{,R})$
and $\eta\simeq1/2$ for $(f_{,RR}/f_{,R})(k/a)^{2}\gg1$, the growth
rate of matter perturbations is larger than that in the $\Lambda$CDM
model, e.g., $\delta_{m}\propto t^{(\sqrt{33}-1)/6}$ during the matter-dominated
epoch \cite{fRStar,TsujifR}.

Substituting Eq.~(\ref{geffeta}) into Eq.~(\ref{Phieff}) we obtain
$\Phi_{{\rm eff}}\simeq-(3/2)(aH/k)^{2}\tilde{\Omega}_{m}\delta_{m}$,
where $\tilde{\Omega}_{m}\equiv\Omega_{m}/f_{,R}$. This means that
the anisotropic parameter $\eta$ between the two gravitational potentials
practically compensates the modification induced by the gravitational
coupling $G_{{\rm eff}}$, i.e. $G_{{\rm eff}}\,(1+\eta)/2=G/f_{,R}$.
During the matter dominance ($\tilde{\Omega}_{m}\simeq1$ and $a\propto t^{2/3}$)
the evolution of the effective gravitational potential on sub-horizon
scales is given by $\Phi_{{\rm eff}}\propto t^{(\sqrt{33}-5)/6}$
\cite{TsujifR}.

\subsection{Brans-Dicke theories}

Brans-Dicke theories \cite{Brans} with the field potential $V(\phi)$
correspond to the choice 
\begin{equation}
K=\frac{\Mpl\omega_{{\rm BD}}X}{\phi}-V(\phi)\,,\qquad G_{3}=0=G_{5}\,,\qquad G_{4}=\frac{1}{2}\Mpl\phi\,,
\end{equation}
 where $\omega_{{\rm BD}}$ is the Brans-Dicke parameter (which is
constant). Compared to original Brans-Dicke theories we have introduced
the reduced Planck mass $\Mpl$ in $K$ and $G_{4}$ such that the
field $\phi$ has a dimension of mass. The difference from $f(R)$
theories appears for the kinetic term $\Mpl\omega_{{\rm BD}}X/\phi$,
in which case $D_{9}=-\Mpl\omega_{{\rm BD}}/\phi$. {}From Eqs.~(\ref{Geff1})
and (\ref{eta1}) it follows that 
\begin{equation}
G_{{\rm eff}}=\frac{\Mpl}{\phi}\frac{4+2\omega_{{\rm BD}}+2(\phi/\Mpl)(Ma/k)^{2}}{3+2\omega_{{\rm BD}}+2(\phi/\Mpl)(Ma/k)^{2}}G\,,\quad\quad\quad\eta=\frac{1+\omega_{{\rm BD}}+(\phi/\Mpl)(Ma/k)^{2}}{2+\omega_{{\rm BD}}+(\phi/\Mpl)(Ma/k)^{2}}\,,\label{GeffBD}
\end{equation}
 where 
\begin{equation}
M^{2}=V_{,\phi\phi}+\frac{\omega_{{\rm BD}}\Mpl}{\phi^{3}}\left[\dot{\phi}^{2}-\phi\left(\ddot{\phi}+3H\dot{\phi}\right)\right]\,.
\end{equation}
 The results (\ref{GefffR}) in $f(R)$ theories can be recovered
by setting $\omega_{{\rm BD}}=0$ in Eq.~(\ref{GeffBD}). It is convenient
to express $M^{2}$ solely in terms of the potential. This can be
done by using the scalar field equation of motion, and we obtain $M^{2}\simeq V_{,\phi\phi}+V_{,\phi}/\phi$,
where we have neglected ${\cal O}(\Mpl H^{2}/\phi)$ terms.

In the limit where $\omega_{{\rm BD}}\to\infty$ or $M^{2}\to\infty$
(with $\phi\simeq M_{{\rm pl}}$) we recover the General Relativistic
behavior: $G_{{\rm eff}}\simeq G$ and $\eta\simeq1$. In the limit
that $M^{2}\to0$ we have $G_{{\rm eff}}\simeq(\Mpl/\phi)(4+2\omega_{{\rm BD}})/(3+2\omega_{{\rm BD}})G$
and $\eta\simeq(1+\omega_{{\rm BD}})/(2+\omega_{{\rm BD}})$ \cite{Tsuji1,Verde}.
The effective gravitational coupling in the latter case also agrees
with the one corresponding to the gravitational force between two
test particles \cite{fujii}.

\subsection{Kinetic gravity braidings}

Let us consider the kinetic gravity braidings \cite{braiding} described
by the Lagrangian 
\begin{equation}
K=K(\phi,X)\,,\qquad G_{3}=G_{3}(\phi,X)\,,\qquad G_{4}=\frac{1}{2}\Mpl^{2}\,,\qquad G_{5}=0\,.
\end{equation}
 Since $A_{6}=-2XG_{3,X}$, $B_{6}=B_{8}=2\Mpl^{2}$, $B_{7}=0$ in
this case, it follows that 
\begin{equation}
G_{{\rm eff}}=\frac{M^{2}-D_{9}(k/a)^{2}}{M^{2}-(D_{9}+2X^{2}G_{3,X}^{2}/\Mpl^{2})(k/a)^{2}}G\,,\quad\quad\quad\eta=1\,,\label{Geffbra}
\end{equation}
 where 
\begin{equation}
D_{9}=-K_{,X}-2\left(G_{{3,X}}+XG_{{3,{\it XX}}}\right)\ddot{\phi}-4HG_{{3,X}}\dot{\phi}+2G_{{3,\phi}}-2XG_{{3,\phi X}}\,.
\end{equation}
 In the limit that $M^{2}\to\infty$ we have $G_{{\rm eff}}\to G$,
so that the General Relativistic behavior is recovered.

Let us consider the theories in which both $K$ and $G_{3}$ depend
only on $X$, i.e. $K=K(X)$ and $G_{3}=G_{3}(X)$. Since $M^{2}=0$,
$G_{3,\phi}=0$, and $G_{{3,\phi X}}=0$ in such theories, the effective
gravitational coupling is given by 
\begin{equation}
G_{{\rm eff}}=G\left\{ 1+\frac{G_{3,X}^{2}\dot{\phi}^{4}}{2\Mpl^{2}[K_{3,X}+2(\ddot{\phi}+2H\dot{\phi})G_{3,X}+G_{3,XX}\dot{\phi}^{2}\ddot{\phi}]-G_{3,X}^{2}\dot{\phi}^{4}}\right\} \,.
\end{equation}
 This result agrees with that derived in Ref.~\cite{Kimura} in which
the authors studied the evolution of perturbations for the functions
$K=-X$ and $G_{3}\propto X^{n}$ (which corresponds to the Dvali
and Turner model \cite{Turner}).

One can also extend the analysis to the case where $G_{4}$ is a function
of $\phi$, i.e. ${\cal L}=G_{4}(\phi)R+K(\phi,X)-G_{3}(\phi,X)\square\phi$.
The perturbation equations for the theories with $G_{3}(\phi,X)=\xi(\phi)X$
were derived in Ref.~\cite{Mukoh} (see also Refs.~\cite{Silva,Koba10}
for specific choices of the functions $G_{4}(\phi)$ and $K(\phi,X)$).

\subsection{Covariant Galileon}

The covariant Galileon without the field potential corresponds to
\cite{DEV} 
\begin{equation}
K=-c_{2}X\,,\qquad G_{3}=\frac{c_{3}}{M^{3}}\, X\,,\qquad G_{4}=\frac{1}{2}\Mpl^{2}-\frac{c_{4}}{M^{6}}\, X^{2}\,,\qquad G_{5}=\frac{3c_{5}}{M^{9}}\, X^{2}\,,\label{Gali}
\end{equation}
 where $c_{i}$ ($i=2,3,4,5$) are dimensionless constants and $M$
is the constant having the dimension of mass. For the choice (\ref{Gali})
we confirmed that all the coefficients in Eqs.~(\ref{Ai})-(\ref{masss})
are equivalent to those given in Ref.~\cite{kase}. Hence Eqs.~(\ref{Geff1})
and (\ref{eta1}) reproduce the effective gravitational coupling $G_{{\rm eff}}$
and the anisotropic parameter $\eta$ derived in \cite{kase}.

For the covariant Galileon there exists a stable de Sitter solution
where $X=$\,constant \cite{cgadark} (see also Refs.~\cite{Galileoninf,Galileoninf2}
for related works). Since $G_{{\rm eff}}$ is larger than $G$ before
the solution reaches the de Sitter attractor, the growth rate of matter
perturbations is larger than that in the $\Lambda$CDM model. In addition
the variation of the effective gravitational potential $\Phi_{{\rm eff}}$
can be more significant than that in $f(R)$ gravity, because $\eta$
can be larger than 1 in the early cosmological epoch \cite{kase}.

\subsection{Field derivative couplings with the Einstein tensor}

The dark energy model of Gubitosi and Linder \cite{Linder} corresponds
to 
\begin{equation}
K=X\,,\qquad G_{4}=\frac{1}{2}\Mpl^{2}\,,\qquad G_{5}=-\lambda\frac{\phi}{\Mpl^{2}}\,,
\end{equation}
 where $\lambda$ is a dimensionless constant (see also Refs.~\cite{Amendola,Germani}).
The Lagrangian in Ref.~\cite{Linder} involves the term $(\lambda/\Mpl^{2})G_{\mu\nu}\nabla^{\mu}\phi\nabla^{\nu}\phi$,
but this is equivalent to $-\lambda(\phi/\Mpl^{2})G_{\mu\nu}\nabla^{\mu}\nabla^{\nu}\phi$
after the integration by parts.

Since $A_{6}=-4\lambda H\dot{\phi}\Mpl^{2}$, $B_{6}=2\Mpl^{2}+4\lambda X/\Mpl^{2}$,
$B_{7}=-4\lambda(\ddot{\phi}+H\dot{\phi})/\Mpl^{2}$, $B_{8}=2\Mpl^{2}-4\lambda X/\Mpl^{2}$,
and $D_{9}=-1-6\lambda H^{2}/\Mpl^{2}-4\lambda\dot{H}/\Mpl^{2}$,
it follows that 
\begin{eqnarray}
G_{{\rm eff}} & = & \left(1+\frac{\xi_{1}}{\xi_{2}}\right)G\,,\label{Geff0}\\
\eta & = & 1-\frac{2\lambda[\Mpl^{2}\dot{\phi}^{2}+2\lambda(2H\dot{\phi}\ddot{\phi}+2\ddot{\phi}^{2}+3H^{2}\dot{\phi}^{2}+2\dot{H}\dot{\phi}^{2})]}{\Mpl^{6}+\lambda\Mpl^{2}(6\Mpl^{2}H^{2}+4\Mpl^{2}\dot{H}+\dot{\phi}^{2})+2\lambda^{2}(7H^{2}\dot{\phi}^{2}+2\dot{H}\dot{\phi}^{2}+4\ddot{\phi}^{2}+8H\dot{\phi}\ddot{\phi})}\,,\label{GeffL}
\end{eqnarray}
 where 
\begin{eqnarray}
\xi_{1} & \equiv & \lambda[3\Mpl^{6}\dot{\phi}^{2}+\lambda\Mpl^{2}(18\Mpl^{2}H^{2}\dot{\phi}^{2}+12\Mpl^{2}\dot{H}\dot{\phi}^{2}+8\Mpl^{2}\ddot{\phi}^{2}-\dot{\phi}^{4})+2\lambda^{2}\dot{\phi}^{3}(8H\ddot{\phi}+9H^{2}\dot{\phi}-2\dot{H}\dot{\phi})]\,,\\
\xi_{2} & \equiv & \Mpl^{10}+2\lambda\Mpl^{6}(3\Mpl^{2}H^{2}+2\Mpl^{2}\dot{H}-\dot{\phi}^{2})+\lambda^{2}\Mpl^{2}(\dot{\phi}^{4}+16\Mpl^{2}H\dot{\phi}\ddot{\phi}-4\Mpl^{2}H^{2}\dot{\phi}^{2}-8\Mpl^{2}\dot{H}\dot{\phi}^{2})\nonumber \\
 &  & +2\lambda^{3}\dot{\phi}^{3}(2\dot{H}\dot{\phi}-8H\ddot{\phi}-9H^{2}\dot{\phi})\,.
\end{eqnarray}
 For $\lambda\neq0$ one has $G_{{\rm eff}}\neq G$ and $\eta\neq1$,
so that the evolution of perturbations is different from that in GR.
Note that the results (\ref{Geff0}) and (\ref{GeffL}) are derived
for the first time in this paper.

\section{Conclusions}

\label{consec} In this paper we have derived the full equations of
scalar density perturbations for the perturbed metric (\ref{permet}).
The Newtonian gauge corresponds to the choice $\chi=0$, which fixes
the temporal part of the gauge-transformation vector. Since the different
gauge choices (such as $\delta\phi=0$) are possible, our linear perturbation
equations can be applied to other gauges as well.

For the perturbations deep inside the Hubble radius the quasi-static
approximation employed in Sec.~\ref{subhsec} is accurate for an
effectively massless scalar field. In fact this is the case for kinetic
gravity braidings and covariant Galileons. In the dark energy models
based on $f(R)$ theories and Brans-Dicke theories, the mass of the
scalar field degree of freedom needs to be large in the region of
high density for consistency with local gravity constraints. In order
to accommodate such cases we have taken into account the effective
mass $M$ in estimating the effective gravitational coupling $G_{{\rm eff}}$.
For the quasi-static approximation to work it is necessary that 
the time-derivatives of the field perturbation $\delta \phi$ are 
neglected relative to the terms including $c_s^2(k^2/a^2)\delta \phi$, 
where $c_s$ is the scalar propagation speed whose explicit expression 
is given in Ref.~\cite{newpaper}.
This implies that the perturbation field cannot be fast oscillating. 
In other words, provided that the oscillating mode of the field perturbation 
is suppressed relative to the matter-induced mode, 
the quasi-static approximation on sub-horizon scales can be trustable. 

In order to estimate the growth rate of perturbations relevant to
large-scale structure and weak lensing, it is sufficient to use the
approximate results of (\ref{Geff1}) and (\ref{Phieff}) with the
anisotropic parameter $\eta$ given by Eq.~(\ref{eta1}). We applied
our formulas to a number of modified gravitational models of dark
energy and found that they nicely reproduce the previously known results.
There are some new models--such as the field derivative couplings
with the Einstein tensor--in which the evolution of perturbations
deserves for further detailed investigation.

For the large-scale perturbations associated with the integrated-Sachs-Wolfe
effect in the CMB anisotropies the sub-horizon approximation in Sec.~\ref{subhsec}
is no longer valid. Instead we need to integrate the perturbation
equations (\ref{eq:Psi})-(\ref{eq:mat2}) numerically, along the
lines of Ref.~\cite{kase}. It will be of interest how the joint
data analysis of CMB combined with the observations of large-scale
structure and weak lensing place constraints on each modified gravitational
model of dark energy accommodated by the general action (\ref{action}).


\section*{ACKNOWLEDGEMENTS}

\label{acknow} A.\,D.\,F.\ and S.\,T.\ were supported by the
Grant-in-Aid for Scientific Research Fund of the JSPS Nos.~10271
and 30318802. T.\,K.\ was supported by JSPS Grant-in-Aid for Research
Activity Start-up No.~22840011. S.\,T.\ also thanks financial support
for the Grant-in-Aid for Scientific Research on Innovative Areas (No.~21111006).
We would like to thank the organizers of the workshop {\em Summer
Institute 2011 (Cosmology \& String)}, where this collaboration was
initiated. 


\appendix

\section{Coefficients relevant to sub-horizon perturbations}

Here we write the explicit forms of the coefficients $A_{6}$, $B_{6}$,
$B_{7}$, $B_{8}$, and $D_{9}$: 
\begin{eqnarray}
A_{6} & = & -2XG_{{3,X}}-4H\left(G_{{4,X}}+2XG_{{4,{\it XX}}}\right)\dot{\phi}+2G_{{4,\phi}}+4XG_{{4,\phi X}}\nonumber \\
 &  & +4H\left(G_{{5,\phi}}+XG_{{5,\phi X}}\right)\dot{\phi}-2{H}^{2}X\left(3G_{{5,X}}+2XG_{{5,{\it XX}}}\right)\,,\\
B_{6} & = & 4[G_{4}-X(\ddot{\phi}\, G_{5,X}+G_{5,\phi})]\,,\\
B_{7} & = & -4G_{{4,X}}H\dot{\phi}-4(G_{{4,X}}+2XG_{{4,{\it XX}}})\ddot{\phi}+4\, G_{{4,\phi}}-8XG_{{4,\phi X}}\nonumber \\
 &  & +4(G_{{5,\phi}}+XG_{{5,\phi X}})\ddot{\phi}-4H[(G_{{5,X}}+XG_{{5,{\it XX}}})\ddot{\phi}-G_{{5,\phi}}+XG_{{5,\phi X}}]\dot{\phi}+4X[G_{{5,\phi\phi}}-(H^{2}+\dot{H})G_{{5,X}}],\\
B_{8} & = & 4[G_{4}-2XG_{4,X}-X(H\dot{\phi}\, G_{5,X}-G_{5,\phi})]\,,\\
D_{9} & = & -K_{,X}-2\left(G_{{3,X}}+XG_{{3,{\it XX}}}\right)\ddot{\phi}-4HG_{{3,X}}\dot{\phi}+2G_{{3,\phi}}-2XG_{{3,\phi X}}\nonumber \\
 &  & +[-4\, H(3\, G_{{4,{\it XX}}}+2XG_{{4,{\it XXX}}})\ddot{\phi}+4H(3G_{{4,\phi X}}-2XG_{{4,\phi{\it XX}}})]\dot{\phi}+(6\, G_{{4,\phi X}}+4XG_{{4,\phi{\it XX}}})\ddot{\phi}\nonumber \\
 &  & -20{H}^{2}XG_{{4,{\it XX}}}+4XG_{{4,\phi\phi X}}-4\dot{H}(G_{4,X}+2XG_{{4,{\it XX}}})-6{H}^{2}G_{4,X}\nonumber \\
 &  & +\{4H(2G_{{5,\phi X}}+XG_{{5,\phi{\it XX}}})\ddot{\phi}-4H[(H^{2}+\dot{H})(G_{{5,X}}+XG_{{5,{\it XX}}})-XG_{{5,\phi\phi X}}]\}\dot{\phi}-4H^{2}X^{2}G_{{5,\phi{\it XX}}}\nonumber \\
 &  & -2H^{2}(G_{{5,X}}+5XG_{{5,{\it XX}}}+2{X}^{2}G_{{5,{\it XXX}}})\ddot{\phi}+2(3H^{2}+2\dot{H})G_{5,\phi}+4\dot{H}XG_{{5,\phi X}}+10{H}^{2}XG_{{5,\phi X}}.
\end{eqnarray}



\end{document}